# *KMT2B*-related disorders: expansion of the phenotypic spectrum and long-term efficacy of deep brain stimulation

Laura Cif,[1,2,†] Diane Demailly,[1,2,†] Jean-Pierre Lin,[3,4,†] Katy E. Barwick,[5] Mario Sa,[3] Lucia Abela,[5] Sony Malhotra,[6] Wui K. Chong,[7] Dora Steel,[5,8] Alba Sanchis-Juan,[9,10] Adeline Ngoh,[5,8] Natalie Trump,[5] Esther Meyer,[5] Xavier Vasques,[11] Julia Rankin,[12] Meredith W. Allain,[13] Carolyn D. Applegate,[14] Sanaz Attaripour Isfahani,[15] Julien Baleine,[16] Bettina Balint,[17,18] Jennifer A. Bassetti,[19] Emma L. Baple,[12,20] Kailash P. Bhatia,[17] Catherine Blanchet,[21] Lydie Burglen,[22] Gilles Cambonie,[16] Emilie Chan Seng,[1,2] Sandra Chantot Bastaraud,[23] Fabienne Cyprien,[1,2] Christine Coubes,[24] Vincent d'Hardemare,[23] Deciphering Developmental Disorders Study,[25] Asif Doja,[26] Nathalie Dorison,[23] Diane Doummar,[27] Marisela E. Dy-Hollins,[28,29] Ellyn Farrelly,[13,30] David R. Fitzpatrick,[31] Conor Fearon,[32] Elizabeth L. Fieg,[33] Brent L. Fogel,[34,35] Eva B. Forman,[36] Rachel G. Fox,[37] Genomics England Research Consortium,[38] William A. Gahl,[39] Serena Galosi,[40] Victoria Gonzalez,[1,2] Tracey D. Graves,[41] Allison Gregory,[37] Mark Hallett,[15] Harutomo Hasegawa,[3,4] Susan J. Hayflick,[37,42] Ada Hamosh,[14] Marie Hully,[43] Sandra Jansen,[44] Suh Young Jeong,[37] Joel B. Krier,[33] Sidney Krystal,[45] Kishore R. Kumar,[46,47,48] Chloé Laurencin,[49] Hane Lee,[35,50] Gaetan Lesca,[51] Laurence Lion François,[52] Timothy Lynch,[32,53] Neil Mahant,[54] Julian A. Martinez-Agosto,[35,55] Christophe Milesi,[16] Kelly A. Mills,[56] Michel Mondain,[21] Hugo Morales-Briceno,[54,57] NIHR BioResource,[9] John R. Ostergaard,[58] Swasti Pal,[59] Juan C. Pallais,[33] Frédérique Pavillard,[60] Pierre-Francois Perrigault,[60] Andrea K. Petersen,[61] Gustavo Polo,[62] Gaetan Poulen,[1,2] Tuula Rinne,[44] Thomas Roujeau,[1] Caleb Rogers,[37] Agathe Roubertie,[63,64] Michelle Sahagian,[65,66] Elise Schaefer,[67] Laila Selim,[68] Richard Selway,[69] Nutan Sharma,[28,29,70] Rebecca Signer,[35] Ariane G. Soldatos,[39] David A. Stevenson,[13] Fiona Stewart,[71] Michel Tchan,[57,72] Undiagnosed Diseases Network,[39] Ishwar C. Verma,[59] Bert B. A. de Vries,[44] Jenny L. Wilson,[73] Derek A. Wong,[55] Raghda Zaitoun,[74] Dolly Zhen,[37] Anna Znaczko,[71] Russell C. Dale,[75,76] Claudio M. de Gusmão,[29,77] Jennifer Friedman,[65,66,78,79] Victor S. C. Fung,[54,57] Mary D. King,[36,53] Shekeeb S. Mohammad,[75,76] Luis Rohena,[80,81] Jeff L. Waugh,[82] Camilo Toro,[39] F. Lucy Raymond,[9,83] Maya Topf,[6] Philippe Coubes,[1,2,‡] Kathleen M. Gorman,[5,8,‡] and Manju A. Kurian[5,8,‡]

[†,‡]These authors contributed equally to this work.

**Heterozygous mutations in *KMT2B* are associated with an early-onset, progressive and often complex dystonia (DYT28). Key characteristics of typical disease include focal motor features at disease presentation, evolving through a caudocranial pattern into generalized dystonia, with prominent oromandibular, laryngeal and cervical involvement. Although *KMT2B*-related disease is emerging as one of the most common causes of early-onset genetic dystonia, much remains to be understood about the full spectrum of the disease. We describe a cohort of 53 patients with *KMT2B* mutations, with detailed delineation of their clinical phenotype and molecular genetic features. We report new disease presentations, including atypical patterns of dystonia evolution and a subgroup of patients with a non-dystonic neurodevelopmental phenotype. In addition to the previously reported systemic features, our study has identified co-morbidities, including the risk of status dystonicus, intrauterine growth retardation, and endocrinopathies. Analysis of this study cohort ($n$ = 53) in tandem with published cases ($n$ = 80) revealed that patients with chromosomal deletions and protein truncating variants had a significantly higher burden of systemic disease (with earlier onset of dystonia) than those with missense variants. Eighteen individuals had detailed longitudinal data available after insertion of deep brain stimulation for medically refractory dystonia. Median age at deep brain stimulation was 11.5 years (range: 4.5–37.0 years). Follow-up after deep brain stimulation ranged from 0.25 to 22 years. Significant improvement of motor function and disability (as assessed by the Burke Fahn Marsden's Dystonia Rating Scales, BFMDRS-M and BFMDRS-D) was evident at 6 months, 1 year and last follow-up (motor, $P$ = 0.001, $P$ = 0.004, and $P$ = 0.012; disability, $P$ = 0.009, $P$ = 0.002 and $P$ = 0.012). At 1 year post-deep brain stimulation, >50% of subjects showed BFMDRS-M and BFMDRS-D improvements of >30%. In the long-term deep brain stimulation cohort (deep brain stimulation inserted for >5 years, $n$ = 8), improvement of >30% was maintained in 5/8 and 3/8 subjects for the BFMDRS-M and BFMDRS-D, respectively. The greatest BFMDRS-M improvements were observed for trunk (53.2%) and cervical (50.5%) dystonia, with less clinical impact on laryngeal dystonia. Improvements in gait dystonia decreased from 20.9% at 1 year to 16.2% at last assessment; no patient maintained a fully independent gait. Reduction of BFMDRS-D was maintained for swallowing (52.9%). Five patients developed mild parkinsonism following deep brain stimulation. *KMT2B*-related disease comprises an expanding continuum from infancy to adulthood, with early evidence of genotype-phenotype correlations. Except for laryngeal dysphonia, deep brain stimulation provides a significant improvement in quality of life and function with sustained clinical benefit depending on symptoms distribution.**

1 Département de Neurochirurgie, Unité des Pathologies Cérébrales Résistantes, Unité de Recherche sur les Comportements et Mouvements Anormaux, Hôpital Gui de Chauliac, Centre Hospitalier Régional Montpellier, Montpellier, France
2 Faculté de médecine, Université de Montpellier, France
3 Complex Motor Disorder Service, Children's Neurosciences Department, Evelina London Children's Hospital, Guy's and St Thomas' NHS Foundation Trust, London, UK
4 Children's Neuromodulation Group, Women and Children's Health Institute, Faculty of life Sciences and Medicine (FOLSM), King's Health Partners, London, UK
5 Molecular Neurosciences, Developmental Neurosciences, UCL Great Ormond Street Institute of Child Health, London, UK
6 Institute of Structural and Molecular Biology, Department of Biological Sciences, Birkbeck College, University of London, London, UK
7 Developmental Imaging and Biophysics, UCL Great Ormond Street Institute of Child Health, London, UK
8 Department of Neurology, Great Ormond Street Hospital, London, UK
9 NIHR BioResource, Cambridge University Hospitals NHS Foundation Trust, Cambridge, UK
10 Department of Haematology, NHS Blood and Transplant Centre, University of Cambridge, Cambridge, UK
11 European IBM Systems Center, Montpellier, France
12 Clinical Genetics, Royal Devon and Exeter NHS Foundation Trust, Exeter, UK
13 Division of Medical Genetics, Department of Pediatrics, Stanford University, Palo Alto, CA, USA
14 McKusick-Nathans Institute of Genetic Medicine, Johns Hopkins University School of Medicine, Baltimore, MD, USA
15 Human Motor Control Section, National Institute of Neurological Disorders and Stroke, National Institutes of Health, Bethesda, MD, USA
16 Unité de Soins Intensifs et Réanimation Pédiatrique et Néonatale, Hôpital Universitaire de Montpellier, Montpellier, France
17 Department of Clinical and Movement Neurosciences, UCL Queen Square Institute of Neurology, London, UK
18 Department of Neurology, University Hospital Heidelberg, Heidelberg, Germany
19 Division of Medical Genetics, Department of Pediatrics, Weill Cornell Medical College, New York, NY, USA
20 Institute of Biomedical and Clinical Science RILD Wellcome Wolfson Centre, University of Exeter Medical School, Royal Devon and Exeter NHS Foundation Trust, Exeter, UK
21 Département d'Oto-Rhino-Laryngologie et Chirurgie Cervico-Faciale, Hôpital Universitaire de Montpellier, Montpellier, France
22 Département de génétique médicale, APHP Hôpital Armand Trousseau, Paris, France
23 Unité Dyspa, Neurochirurgie Pédiatrique, Hôpital Fondation Rothschild, Paris, France
24 Département de Génétique médicale, Maladies rares et médecine personnalisée, CHU Montpellier, Montpellier, France
25 DDD Study, Wellcome Trust Sanger Institute, Hinxton, Cambridge, UK
26 Division of Neurology, Children's Hospital of Eastern Ontario, Ottawa, ON, Canada

27 Neuropédiatrie, Centre de référence neurogénétique mouvement anormaux de l'enfant, Hôpital Armand Trousseau, AP-HP, Sorbonne Université, France
28 Department of Neurology, Massachusetts General Hospital, Boston, MA, USA
29 Department of Neurology, Harvard Medical School, Boston, MA, USA
30 Department of Pediatrics, Lucile Packard Children's Hospital at Stanford, CA, USA
31 Human Genetics Unit, Medical and Developmental Genetics, University of Edinburgh Western General Hospital, Edinburgh, Scotland, UK
32 Department of Neurology, The Dublin Neurological Institute at the Mater Misericordiae University Hospital, Dublin, Ireland
33 Division of Genetics, Department of Medicine, Brigham and Women's Hospital, Harvard Medical School, Boston, MA, USA
34 Department of Neurology, David Geffen School of Medicine, University of California, Los Angeles, CA, USA
35 Department of Human Genetics, David Geffen School of Medicine, University of California, Los Angeles, CA, USA
36 Department of Paediatric Neurology and Clinical Neurophysiology, Children's Health Ireland at Temple Street, Dublin, Ireland
37 Department of Molecular and Medical Genetics, Oregon Health and Science University, Portland, OR, USA
38 Genomics England, London, UK
39 Undiagnosed Diseases Program, National Human Genome Research Institute, National Institutes of Health, Bethesda, MD, USA
40 Department of Human Neuroscience, Sapienza University of Rome, Rome, Italy
41 Department of Neurology, Hinchingbrooke Hospital, North West Anglia NHS Foundation Trust, Huntingdon, UK
42 Department of Paediatrics, Oregon Health and Science University, Portland, OR, USA
43 Département de Neurologie, APHP-Necker-Enfants Malades, Paris, France
44 Department of Human Genetics, Donders Institute for Brain, Cognition and Behaviour, Radboud University Medical Center, Nijmegen, The Netherlands
45 Département de Neuroradiologie, Hôpital Fondation Rothschild, Paris
46 Translational Genomics Group, Kinghorn Centre for Clinical Genomics, Garvan Institute of Medical Research, Darlinghurst, NSW, Australia
47 Department of Neurogenetics, Kolling Institute, University of Sydney and Royal North Shore Hospital, St Leonards, NSW, Australia
48 Molecular Medicine Laboratory, Concord Hospital, Sydney, NSW, Australia
49 Département de Neurologie, Hôpital Neurologique Pierre Wertheimer, Lyon, France
50 Department of Pathology and Laboratory Medicine, David Geffen School of Medicine, University of California, Los Angeles, CA, USA
51 Département de Génétique, Hôpital Universitaire de Lyon, Lyon, France
52 Département de Pédiatrie, Hôpital Lyon-Sud, Pierre-Bénite, France
53 UCD School of Medicine and Medical Science, University College Dublin, Dublin, Ireland
54 Movement Disorders Unit, Department of Neurology, Westmead Hospital, Westmead, NSW, Australia
55 Division of Medical Genetics, Department of Pediatrics, David Geffen School of Medicine, University of California, Los Angeles, CA, USA
56 Department of Neurology, Johns Hopkins University School of Medicine, Baltimore, MD, USA
57 Sydney Medical School, University of Sydney, Sydney, NSW, Australia
58 Centre for Rare Diseases, Aarhus University Hospital, Aarhus, Denmark
59 Institute of Genetics and Genomics, Sir Ganga Ram Hospital, Rajender Nagar, New Delhi, India
60 Département d'Anesthésie-Réanimation Gui de Chauliac, Centre Hospitalier Universitaire de Montpellier, Montpellier, France
61 Randall Children's Hospital, Portland, OR, USA
62 Département de Neurochirurgie Fonctionnelle, Hôpital Neurologique et Neurochirurgical, Pierre Wertheimer, Lyon, France
63 Département de Neuropédiatrie, Hôpital Universitaire de Montpellier, Montpellier, France
64 INSERM U1051, Institut des Neurosciences de Montpellier, Montpellier, France
65 Division of Neurology, Rady Children's Hospital San Diego, CA, USA
66 Department of Neuroscience, University of California San Diego, CA, USA
67 Medical Genetics, Hôpitaux Universitaires de Strasbourg, Strasbourg, France
68 Cairo University Children Hospital, Pediatric Neurology and Metabolic division, Cairo, Egypt
69 Department of Neurosurgery, King's College Hospital, London, UK
70 Department of Neurology, Brigham and Women's Hospital, Boston, MA, USA
71 Department of Genetic Medicine, Belfast Health and Social Care Trust, Belfast, UK
72 Department of Genetics, Westmead Hospital, Westmead, NSW, Australia
73 Division of Pediatric Neurology, Department of Pediatrics, Oregon Health and Science University, Portland, OR, USA
74 Department of Paediatrics, Neurology Division, Ain Shams University Hospital, Cairo, Egypt
75 Department of Paediatric Neurology, The Children's Hospital at Westmead, NSW, Australia
76 Faculty of Medicine and Health, Sydney Medical School, University of Sydney, Sydney NSW, Australia
77 Department of Neurology, Boston Children's Hospital, Boston, MA, USA
78 Departments of Paediatrics, University of California, San Diego, CA, USA
79 Rady Children's Institute for Genomic Medicine, San Diego, CA, USA
80 Division of Medical Genetics, Department of Pediatrics, San Antonio Military Medical Center, San Antonio, TX, USA
81 Department of Pediatrics, Long School of Medicine, UT Health, San Antonio, TX, USA

82 Division of Pediatric Neurology, Department of Pediatrics, University of Texas Southwestern, Dallas, TX, USA
83 Department of Medical Genetics, Cambridge Institute for Medical Research, University of Cambridge, Cambridge, UK

Correspondence to: Professor Manju Kurian
UCL Professor of Neurogenetics and NIHR Research Professor
Room 111, Level 1, UCL Great Ormond Street Institute of Child Health
30 Guilford Street, London, WC1N 1EH, UK
E-mail: manju.kurian@ucl.ac.uk

Correspondence may also be addressed to: Dr Laura Cif, MD, PhD
Département de Neurochirurgie, Hôpital Gui de Chauliac, Centre Hospitalier Universitaire
Montpellier, Unité de Recherche sur les Comportements et Mouvements Anormaux
80 Avenue Augustin Fliche, 34000 Montpellier, France
E-mail: a-cif@chu-montpellier.fr

**Abbreviations:** BFMDRS = Burke-Fahn-Marsden Dystonia Rating Scale; DBS = deep brain stimulation; GP(i) = globus pallidus (internus); WES = whole-exome sequencing

# Introduction

Dystonia is defined as a hyperkinetic motor disorder characterized by involuntary and sustained muscle contractions causing abnormal, twisted and often painful movements and postures (Albanese *et al.*, 2013). With the advent of next-generation sequencing, the landscape of genetic dystonia has been revolutionized by the discovery of novel dystonia genes and new gene-associated phenotypes (Lohmann and Klein, 2017). It is increasingly recognized that most genetically-determined dystonias, particularly those of childhood-onset, are characterized by additional neurological, neuropsychiatric and systemic features. These 'complex dystonia' phenotypes can often pose a significant diagnostic challenge for clinicians.

Recently, mutations in the lysine-specific histone methyltransferase 2B gene, *KMT2B*, (hg38: chr19:35717817-35738879, OMIM 606834*)* were identified in individuals with early-onset dystonia (DYT28, DYT-*KMT2B*) (Zech *et al.*, 2016; Meyer *et al.*, 2017). Typically, affected patients initially present with a focal, often lower-limb dystonia that subsequently evolves into generalized dystonia with prominent cranial, cervical and laryngeal involvement. In many patients, additional clinical features have also been reported, including dysmorphism, short stature, intellectual disability, eye movement abnormalities and psychiatric comorbidities. Although *KMT2B* was only recently identified, more than 80 patients are already published (Zech *et al.*, 2016, 2017*a*, *b*; Lange *et al.*, 2017; Meyer *et al.*, 2017; Reuter *et al.*, 2017; Baizabal-Carvallo and Alonso-Juarez, 2018; Faundes *et al.*, 2018; Hackenberg *et al.*, 2018; Kawarai *et al.*, 2018; Zhao *et al.*, 2018; Brás *et al.*, 2019; Carecchio *et al.*, 2019; Dafsari *et al.*, 2019; Dai *et al.*, 2019*a*, *b*; Klein *et al.*, 2019; Kumar *et al.*, 2019; Ma *et al.*, 2019; Zhou *et al.*, 2019; Cao *et al.*, 2020; Miyata *et al.*, 2020; Mun *et al.*, 2020) rendering *KMT2B*-dystonia an emerging key player in childhood-onset genetic dystonia, accounting for an estimated 21.5% of cases (Carecchio *et al.*, 2019).

Since 1996, deep brain stimulation (DBS) has been proposed as a treatment of severe childhood-onset dystonia, with clinical outcome dependent on a number of factors, including underlying aetiology, patient age and disease severity at the time of surgery (Coubes *et al.*, 1999, 2000; Cif *et al.*, 2010; Gruber *et al.*, 2010; Panov *et al.*, 2012, 2013; Lumsden *et al.*, 2013; Krause *et al.*, 2015; Koy *et al.*, 2018). Despite extensive investigations, a significant number of dystonia cases remain undiagnosed, though symptomatic treatment with DBS is nevertheless employed (Jinnah *et al.*, 2017). With the advance of next generation sequencing technologies and the identification of distinct genetic dystonia syndromes, accurate DBS prognostication is increasingly possible based on underlying aetiology (Coubes *et al.*, 2000; Cif *et al.*, 2010; Gruber *et al.*, 2010; Timmermann *et al.*, 2010; Panov *et al.*, 2012; Krause *et al.*, 2015; Koy *et al.*, 2018). However, little is known about the long-term outcome with DBS in *KMT2B*-dystonia, though short-term benefit has been reported (Zech *et al.*, 2016, 2017*a*; Meyer *et al.*, 2017; Kawarai *et al.*, 2018; Carecchio *et al.*, 2019; Dafsari *et al.*, 2019; Kumar *et al.*, 2019; Cao *et al.*, 2020; Miyata *et al.*, 2020; Mun *et al.*, 2020).

In this study, we report 53 individuals (study cohort) with either *KMT2B* intragenic variants or chromosomal microdeletions encompassing this gene. This report provides a deeper understanding of the spectrum of the *KMT2B*-related phenotype, identifying new clinical features and a distinct group of *KMT2B* patients presenting with a neurodevelopmental disorder in the absence of dystonia, a likely under-reported *KMT2B* phenotype. Furthermore, we have analysed the features of our study cohort ($n$ = 53) together with previously published cases ($n$ = 80). Review of this extended cohort ($n$ = 133) has enabled us to better delineate the spectrum of clinical symptoms, determine the incidence of dystonia-associated phenotypes, and further understand the types of mutations observed in *KMT2B*-related disease. In addition, where sufficient data were available, we report on the outcome (up to 22 years) with DBS in 18 patients (DBS subcohort); this is the largest DBS cohort reported to date, and the data will aid clinicians in counselling patients and families.

# Materials and methods

## Ethical approvals

This study was approved by the National Research Ethics Services in the UK (IRAS project ID: 248447), Great Ormond Street Hospital Research Management and Governance Team (18NM21) and Internal Review Board of Montpellier University Hospital (Ethics Board number 2018_IRB-MTP_11-11). Written informed consent was obtained for all participants in whom research genetic testing was undertaken and for publication of photographs and videos (Supplementary Table 1).

## Patient ascertainment and review of clinical features

Through collaboration with 23 international centres (Supplementary Table 1), 53 patients with disease-associated mutations in *KMT2B* and chromosomal microdeletions including *KMT2B* were ascertained for inclusion in this study (study cohort). For each confirmed case, each study centre completed a case note review and returned anonymized data through a standardized study proforma. Neuroimaging was performed as part of routine clinical care in 48 patients; imaging was available for review in 21 cases, undertaken by a paediatric neuroradiologist (W.K.C.).

DBS data were collected for 18 subjects who underwent DBS (DBS subcohort) to the globus pallidus pars interna (GPi) for severe and medically refractory dystonia. Twelve patients (Patients 1, 9, 10, 17, 19–21, 26, 31–33 and 37; Table 1 and Supplementary material) are in the study cohort and six patients (Patients 53–58; Supplementary Table 7) were previously reported but either before DBS or without detailed information about DBS setting and long-term outcome (Meyer *et al.*, 2017). Three individuals were described previously before genetic diagnosis (Patients 9, 10 and 32) (Coubes *et al.*, 1999; Nerrant *et al.*, 2018). For the DBS subcohort, each study centre completed a case note review and returned anonymized data through a second standardized proforma. Long-term follow-up group with DBS was defined as follow-up of at least 5 years (Volkmann *et al.*, 2012; Cif *et al.*, 2013). Dystonia severity and related disability at baseline and during follow-up with DBS was assessed by the motor and disability sections of the Burke-Fahn-Marsden Dystonia Rating Scale (BFMDRS) (Burke *et al.*, 1985). Suboptimal response to DBS was defined as less than 30% reduction in the BFMDRS-M compared to baseline (Pauls *et al.*, 2017).

## Molecular genetic investigations and *in silico* modelling

Mutations were identified through a variety of different methods, including microarray ($n = 2$) research Sanger sequencing ($n = 10$), diagnostic gene panels for dystonia ($n = 7$), diagnostic exome/genome analysis ($n = 22$) and research exome/genome sequencing ($n = 12$) (Tables 1, 3 and Supplementary material).

### Research Sanger sequencing confirmation

For Patients 8–10, 17, 20, 21, 32, 37, 39 and 46, direct Sanger sequencing was carried out to (i) screen the entire coding region of the *KMT2B* gene from the study cohort; (ii) confirm *KMT2B* variants identified by research whole-exome or whole-genome sequencing; and (iii) to establish familial segregation. *KMT2B* wild-type sequence was obtained from Ensembl (ENSG00000272333, transcript ENST00000222270) and primers were designed for all 37 exon–intron boundaries using online Primer3Plus software. Purification of PCR products was undertaken using MicroCLEAN (Clent Life Science) and sequencing with the BigDye™ Terminator v1.1 Cycle Sequencing Kit (Applied Biosystems™). Sequencing reactions were run on an ABI PRISM 3730 DNA Analyzer (Applied Biosystems™) and analysed using Mutation Surveyor® software.

### Research whole-exome and whole-genome sequencing

A summary of the different methods used for research whole-exome and genome sequencing performed in different laboratories is provided in the Supplementary material.

### Determination of mutation pathogenicity

Identified *KMT2B* variants were analysed to determine whether they were previously described in other patients, reported on mutation databases (ClinVar) or novel. Protein-truncating variants included nonsense variants, intragenic deletions and duplications (which are predicted to truncate the C-terminus of the protein), one in-frame deletion (that, although is not predicted to truncate the C-terminus of the protein, does shorten the protein), and splice site changes predicted to cause aberrant splicing (Supplementary Table 2), classified pathogenic as per American College of Medical Genetics and Genomics (ACMG) guidelines (Richards *et al.*, 2015). Missense substitutions were suggested to be disease-causing if the Combined Annotation Dependent Depletion (CADD) was $>20$ (v 1.4), absent in Genome Aggregation Database (gnomAD) and the variant was predicted to be disease-causing by at least two *in silico* prediction programs (PolyPhen-2, SIFT, Provean and MutationTaster) with confirmation of pathogenicity likelihood, as defined by ACMG guidelines (Richards *et al.*, 2015) (Supplementary Table 3).

### Missense constraint analysis

Constraint analysis was performed on all reported pathogenic variants in the extended cohort (study cohort and published cases) and compared to reported variants in gnomAD (Karczewski *et al.*, 2020).

### Protein structure-function *in silico* modelling

Homology modelling was undertaken as previously (Meyer *et al.*, 2017) for the PHD-like and SET-binding domains of KMT2B (NP_055542.1). Variants were analysed for a change in free energy using SDM2 and mCSM (Pires *et al.*, 2014; Pandurangan *et al.*, 2017).The predicted negative ddG values imply that the mutation destabilises the protein structure whereas the positive ddG predicts stabilization of protein structure upon mutation.

**Table 1 Phenotypic characteristics of the movement disorder in patients within the study cohort with *KMT2B*-related dystonia (*n* = 44)**

| Pt | Variant<br>Inheritance<br>How diagnosed | Age, y: Sex | AOO motor, y | Motor and other features at onset | Bilateral LL/UL age, y | Cranial / cervical / caryngeal age, y | Symptoms of cranial, cervical, laryngeal dystonia | MRI age, yGPi hypo. Additional features | Medications: response | GPI-DBS age, y |
|---|---|---|---|---|---|---|---|---|---|---|
| 1[a] | Deletion: chr19:34521462–3819173<br>*De novo*<br>Microarray | F: 5 | 2.5 | Unilateral LL dystonia | 3.5/NR | NR/ND/4.25 | Dysarthria, dysphonia, swallowing difficulties | 4.9<br>Y | BZD, L-DOPA, THY: no benefit | Yes[c]<br>4.5 |
| 2 | Deletion: chr19:3556008–35880945<br>Unknown<br>Microarray | 6: M | 3 | Bilateral; LL dystonia | 3.25/4.75 | 5.75/NR/ND | Dysarthria→anarthria; mild rotary torticollis | 4.5<br>N | NR | No |
| 3 | c.12_24dup13 p.Ser9Glyfs*111<br>*de novo*<br>Research WGS | 22: F | 5 | Unilateral LL dystonia | 6/11 | 15/17/18 | Dysarthria→anarthria; dysphagia; retrocollis; tridor | 8,9,11,13,17<br>Y[b]<br>Less evident aged 17 compared to 13 | BFN, CBZ, CPM, DPM, ITBFN, L-DOPA, THY: no benefit; TIZ: clinical benefit | Yes<br>22 |
| 4 | c.188delG<br>p.Ala40Profs*6<br>*de novo*<br>Diagnostic WGS<br>+twin of Pt 5 | 19: M | 7-8 | Dysarthria; precocious puberty; deceleration of growth | 14/14–15 | 7–8/13/7–8 | Dysarthria→anarthria; drooling; swallowing difficulties; jaw-opening dystonia; torticollis | 12,16,18<br>Y | BFN, CLO, CPM, TBZ, THY: no benefit; ARI, BTX: clinical benefit | Yes<br>18 |
| 5 | c.188delG<br>p.Ala40Profs*6<br>*de novo*<br>Diagnostic WGS<br>+twin of Pt 4 | 19: M | 7-8 | Dysarthria; precocious puberty | 14/14–15 | 7–8/7–8/7–8 | Dysarthria; drooling; swallowing difficulties; torticollis | 12,13<br>Y | BFN, THY, ARI: no benefit; BTX: clinical benefit | Yes<br>18 |
| 6 | c.816dupC<br>p.Gly273Argfs*61<br>*de novo*<br>Diagnostic WES | 12.75: F | 5–6 | LL dystonia; dysarthria | 6–7/9 | 5.5/ND/ND | Dysarthria; swallowing difficulties | 11<br>Y[b] | NR | No |
| 7 | c.850C > T<br>p.Gln284*<br>*de novo*<br>Diagnostic Panel | 13: F | 6 | Cervical dystonia with febrile illness, followed by unilateral LL dystonia | 6.5/unilateral only 12 | 12/6/ND | Dysarthria | 10<br>Y | L-DOPA: no benefit; CBZ: 10–20% reduction in exercise induced dystonia | No |
| 8 | c.1107dupC<br>p.Glu370Argfs*19<br>*de novo*<br>Research SS | 26: M | 4 | Unilateral LL dystonia | 8/10 | 8/15/10 | Dysarthria→ anarthria; spasmodic dysphonia; swallowing difficulties; torticollis; jaw-opening dystonia | 15.75<br>Y[b] | L-DOPA trial, ROP: no benefit | No |

Patients 1–8 are shown; details for Patients 9–44 are provided in the Supplementary material. NM_014727.2, GRCh38 (Chr 19). AOO motor = age at onset of motor symptoms; ARI = aripiprazole; BFN = baclofen; BTX = Botulinum toxin; BZD = benzodiazepine; CBZ = carbamazepine; CLO = clonidine; CPM = clonazepam; DPM = diazepam; F = female; hypo. = hypointensity; ITBFN = intra-thecal baclofen; L-DOPA = levodopa/carbidopa; LL = lower limb; M = male; N = no; ND = never developed; NR = not recorded; Pt = patient; ROP = ropinirole; SS = sanger sequencing; TBZ = tetrabenazine; THY = trihexyphenidyl; TIZ = tizanidine; UL = upper limb; VPA = sodium valproate; WES = whole-exome sequencing; WGS = whole-genome sequencing; Y = yes.

[a]Included in the DBS cohort. See Table 4 for further details; [b]Reviewed by Neuroradiologist W.K.C; [c]Longitudinal GPI-DBS data are available.

## Literature review and statistical analysis

A comprehensive search of the medical literature (PubMed, Medline) was conducted to identify all English-language papers reporting patients with *KMT2B*-related disorders (Supplementary Table 4). All papers were reviewed to create a published cohort of cases. The study cohort and published cohort of cases were amalgamated into the extended cohort for further review and statistical analysis (Supplementary Tables 5 and 6). Missense variants not meeting the criteria described above were not included in the analysis. Statistical analysis was performed using SPSS v24 with significance set at $P$-value < 0.05. Parametric tests were performed where the data were normally distributed, and non-parametric tests were used if the data were not normally distributed. To compare the evolution of motor and disability scores with DBS, the Wilcoxon signed-rank test was used. Correlations between age at dystonia onset, time to generalization and dystonia severity preoperatively were studied using non-parametric Spearman's rho test. ANOVA was used to study relationships between the type of mutation, dystonia severity at baseline and response to DBS. The XgBoost tree predictor importance model was used to define features important to predict the type of mutation. We used F-score: the higher the F-score, more important the feature in predicting the type of variant. Variants were grouped as protein-truncating variants, chromosomal microdeletions or missense variants.

## Data availability

The authors confirm that the data supporting the findings of this study are available within the article and its Supplementary material.

# Results

## *KMT2B* molecular genetic features

We identified 53 individuals (35 females) with mutations in *KMT2B* (Tables 1–3 and Supplementary material). A variety of mutations were identified, including protein-truncating variants ($n = 40$), missense variants ($n = 11$) and chromosomal microdeletions ($n = 2$) (Fig. 1). Three patients (Patients 18, 23 and 47) were previously reported briefly as part of a cohort whole-genome sequencing (WGS) study (Kumar *et al.*, 2019). Although parental testing to establish inheritance patterns was not possible for all cases, where it had been undertaken for both parents, it was clear that the majority (36/38, 94.7%) had occurred *de novo* (Tables 1, 3 and Supplementary material). Patient 29 inherited the c.4960T>C variant from her symptomatic mother (Patient 30) who was initially thought to have a neuromuscular disorder. Patient 17 inherited the c.3147_3160del from his mother (Patient 46) with no dystonia but short stature, dysmorphism and intellectual disability. Patients 18 and 47 are siblings; their father died before genetic diagnosis was achieved and therefore segregation studies were not possible. Of note, he was also reported to have clinical symptoms suggestive of dystonia.

All identified mutations were novel apart from c.1656dupC (Patient 9), c.2428C>T (Patient 13), c.4789C>T (Patients 25, 50) and c.4847C>T (Patient 27) (Zech *et al.*, 2016, 2017*b*; Hackenberg *et al.*, 2018; Kawarai *et al.*, 2018; Carecchio *et al.*, 2019; Dai *et al.*, 2019*a*; Zhou *et al.*, 2019; Cao *et al.*, 2020). The same variant (c.4789C>T), was identified in two unrelated patients, one with typical *KMT2B*-related dystonia (Patient 25) and another aged 12 years (Patient 50), who had a neurodevelopmental phenotype without dystonia. The c.3325delC was identified in siblings, one with typical *KMT2B*-dystonia and (Patient 18) and his sister with intellectual disability and short stature (Patient 47) (Kumar *et al.*, 2019). The c.188delG variant was identified in a pair of monozygotic twins (Patients 4 and 5).

### Missense constraint analysis

Constraint analysis was performed on all reported pathogenic variants and compared to reported variants in gnomAD (Fig. 1). Although protein-truncating variants are distributed throughout the protein coding sequence, all missense variants deemed likely to be pathogenic lie within constrained regions for missense variation, which are close to key protein domains such as the SET-binding domain, and PHD, PHD-like and FYR regions.

### Protein structure-function *in silico* modelling

To predict the effects of sequence variants on the structure–function properties of *KMT2B* (NP_055542.1), the site-directed mutant protein models were generated using the *swapaa* command in UCSF Chimera (Pettersen *et al.*, 2004) (with Dunbrack backbone-dependent rotamer library and choosing rotamer based on the lowest clash score, highest number of H-bonds and highest rotamer probability) for all the mutants that fall into the region which could be modelled using homology modelling. Evaluation of the impact of mutations for the modelled variants using SDM2 and mCSM suggests a change in the free energy, with a predicted structure destabilizing effect (Supplementary Table 9). The modelled variants (p.Arg1597Trp, p.Ala1616Val, p.Cys1644Phe, p.Cys1654Arg, p.Arg2649Cys) show disruption of key domains with loss of critical interactions/bonds and effect on protein stability (Fig. 2).

## Clinical features of the study cohort

Overall, 44 patients were identified with a dystonia phenotype, and nine patients with a non-dystonia phenotype.

### Dystonia group

We identified 44 individuals (28 females) with dystonia, with a current median age of 16.0 years (range: 3–44 years). The median age of symptom onset was 5.0 years (range: 1.5–29.0 years), with progression to generalized dystonia, over a median period of 2.0 years (range: 0–10.5 years). Generalization occurred within 6 months of first symptoms in 10 patients.

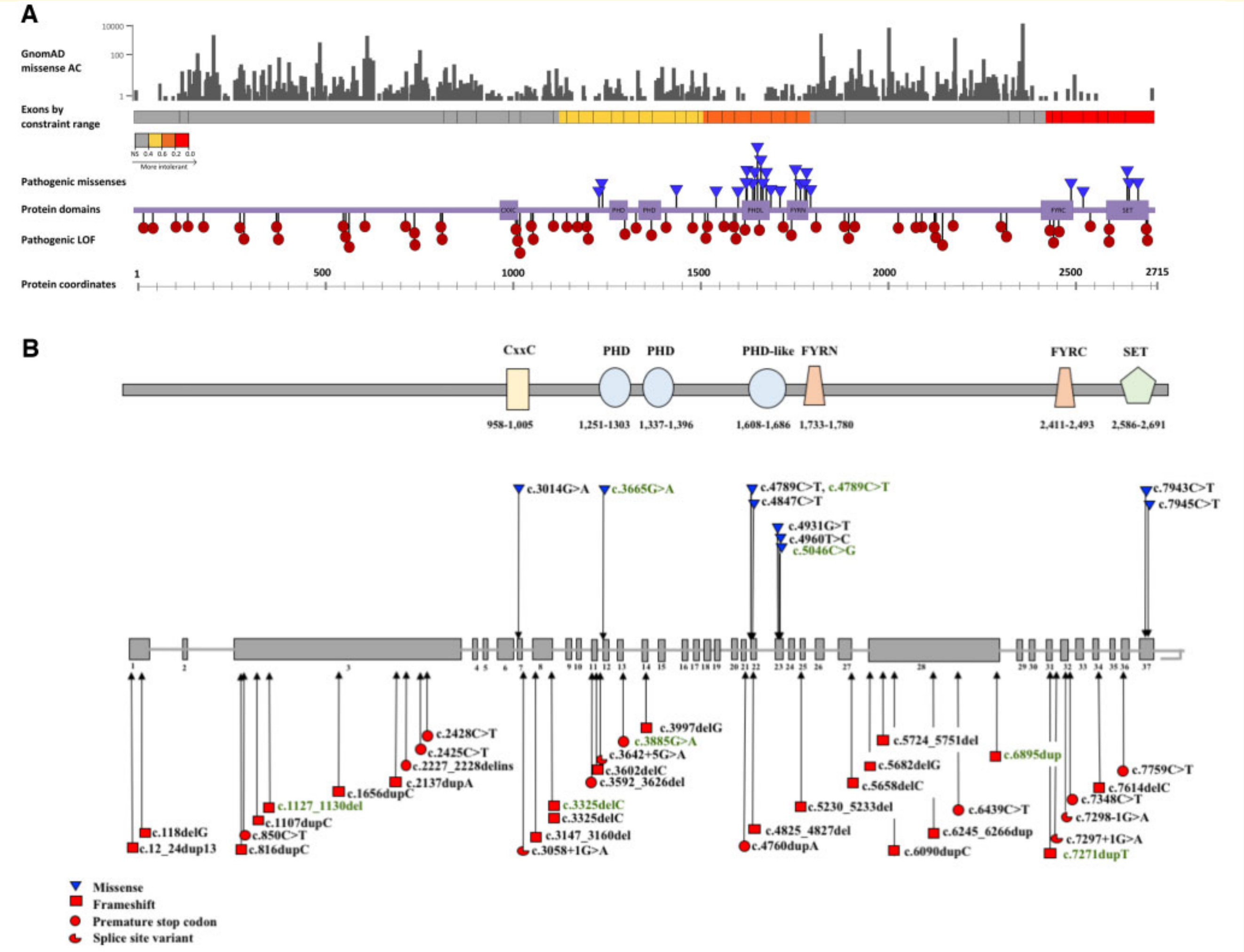

**Figure 1 *KMT2B* missense constraint analysis.** (**A**) Missense allele counts for all *KMT2B* missense variants were obtained from gnomAD v2.1 (Karczewski *et al.*, 2020). All missense amino acid substitutions are represented in grey (*top* section). Exons by constraint ranges were obtained from Decipher v.9.29 (Firth *et al.*, 2009). Schematic representation of the coding exons of *KMT2B* shows the gene regions by predicted intolerance to missense changes, ranging from grey (relatively tolerant) through yellow, orange and red with increasing intolerance (track for exons are coloured by constraint values, obtained from Decipher). Highly constrained regions encompass exons encoding key protein domains, including the CxxC, PHD, PHD-like, FYR and SET-binding domain. All variants reported in the extended cohort are represented. While pathogenic protein truncating variants (red circles) are distributed throughout the protein coding sequence, disease-associated missense variants (blue triangles) appear to localize in regions of missense constraint, which represent key protein domains. Coordinates for the protein domains were obtained from Pfam 32.0 (El-Gebali *et al.*, 2019). (**B**) Schematic representation of *KMT2B* (NM_014727.2) indicating the positions of 22 frameshift insertions and deletions (red squares), 10 stop-gain mutations (red circles), four splice-site variants (red three quarter circles) and nine missense changes (blue inverted triangles) of the study cohort. Mutations associated with dystonic phenotypes are depicted in black and those associated with non-dystonia phenotypes in green. The functional domain architecture of *KMT2B* is located above the gene diagram.

The majority presented initially with lower limbs symptoms (29/44); foot posturing, new-onset toe-walking or gait difficulties (Table 1 and Supplementary material). Atypical first disease presentations included isolated upper limb dystonia or oromandibular dystonia without limb features.

The median age of development of bilateral lower limb dystonia was 6.0 years (range: 1.5–20.0 years), with caudocranial progression of dystonia and bilateral upper limb involvement by a median age of 8.0 years (range: 2.0–18.0 years). Cranial features were evident by a median age of 7.5 years (range: 2.0–23.0 years), laryngeal symptoms by a median age of 9.0 years (range: 3.5–19.0 years) and cervical dystonia from a median age of 9.0 years (range: 2.0–17.0 years). Laryngeal, oromandibular and cervical involvement became a prominent feature of the disease in the majority, often very disabling, requiring enteral feeding to maintain nutrition, assistive technology for communication and adapted seating. Status dystonicus occurred in 11.4% ($n = 5$) of the study cohort. The majority of patients trialled many different anti-dystonia medications with either no clinical benefit or minimal sustained improvement (Table 1 and Supplementary material). In the study cohort, 23 patients had DBS inserted: results are reported in the focused DBS cohort for 12 subjects, for whom longitudinal data of clinical evolution with DBS was available.

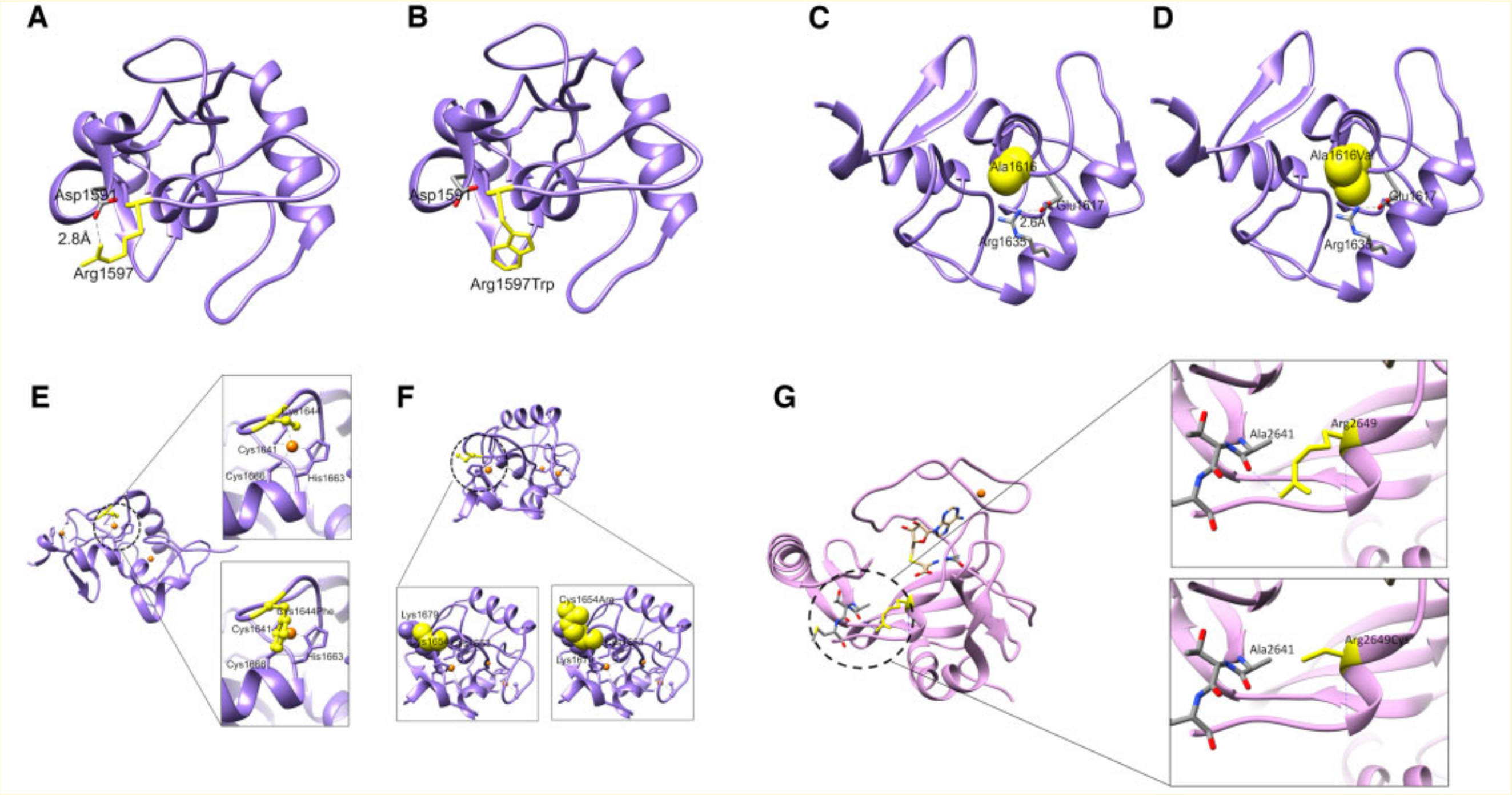

**Figure 2 Predicted effect of KMT2B variants on structure-function properties.** (**A**–**F**) Structural modelling for PHD-like domain of KMT2B (residues 1574-1688). (**A**) The wild-type residue Arg1597 (yellow) lies on the exposed flexible loop, which is involved in the salt bridge with Asp1591. (**B**) Substitution of the arginine with a tryptophan is predicted to remove this salt bridge interaction by placing a bulkier hydrophobic side chain in this position. (**C**) The wild-type residue Ala1616 (yellow) is close to the residue Arg1635, which is involved in salt bridge with Glu1617. (**D**) Ala1616Val may result in a steric clash with Arg1635 and disrupt the salt bridge between Arg1635 and Glu1617. (**E**) This PHD-like domain is known to have three zinc fingers (zinc ions shown in orange). Cys1644 is involved in formation of one of the zinc-fingers (*top*). Substitution with a phenylalanine will result in loss of coordination of zinc ion in a zinc finger and a detrimental effect on the protein function (*bottom*). (**F**) Cys1654 is present adjacent to Cys1653 which is involved in coordinating zinc ion in another zinc finger motif (*bottom left*). Substitution of a cysteine with an arginine might cause a steric clash and repulsion with Lys1679 present in its vicinity, hence impacting on protein structure (*bottom right*). (**G**) Structural modelling for SET domain of KMT2B (residues 2539–2715). The side chain of Arg2649 forms an H-bond with the backbone of Ala2641 (*top*). Substitution at amino acid 2649 of the arginine with a cysteine will disrupt this bond (*bottom*) and impact the stability of the domain.

In our cohort, additional features were present in all (44/44), including short stature (height <2nd centile) (71.1%), microcephaly with occipitofrontal circumference (OFC) <2nd centile (68.6%) and dysmorphism (52.4%) (Table 2, Supplementary Fig. 1 and Supplementary material). Developmental delay preceding the onset of dystonia was reported in 14 children (34.1%). Subsequent cognitive difficulties were noted in 47.6%, ranging from mild to severe intellectual disability. Autism spectrum disorder was reported in six individuals (14.3%). Psychiatric features such as attention deficit hyperactivity disorder and anxiety were identified in 18 (42.9%) cases. Dysmorphic features such as an elongated face, bulbous nasal tip and clinodactyly were present in 22 individuals (52.4%). Endocrinopathies, including hypothyroidism and precocious puberty, were reported in 10 cases (23.3%). Ophthalmological defects (18 cases, 42.9% including refractive errors, end-gaze nystagmus and slow saccades), skin features (three cases, 7.1%) and other systemic features such as cyclical neutropenia, autoimmune hepatitis or IgG deficiency (13 cases, 30.2%) were also reported (Table 2 and Supplementary material).

Neuroimaging was available for review for 21 patients and systematically reviewed by a single neurologist (W.K.C.). Bilateral symmetrical hypointensity of the globus pallidus (GP) with a distinct hypointense lateral streak of GP externa was evident for 17/21 patients (mean age of imaging 11.0 years, range: 4.0–25.0 years) (Supplementary Fig. 3). This was most obvious in susceptibility-weighted images (SWI) and B0 images, with concomitant $T_2$ images normal in some instances (Patients 20 and 32). Those without GP changes on MRI tended to be older, with a mean age 23.6 years at neuroimaging (range: 8.0–36.0 years). Serial scans were available in six cases, showing changes in GP hypointensity over time. For Patient 39, MRI brain scan at 11.8 years showed greater GP hypointensity than at 9 years. For Patient 3, MRI brain scan at 17 years showed a reduction in GP hypointensity compared to neuroimaging at 13 years (Supplementary Fig. 3). Other radiological features identified included non-specific white matter changes (Patients 25 and

**Table 2 Additional features in patients with *KMT2B*-related dystonia ($n$ = 44)**

| Pt | Birth/neonatal issues | DD/pre-dystonia | ID/ASD | Microcephaly/OFC (centile) | Weight/height (centile) | Dysmorphism | Eyes | Dermatology | Psychiatric | Other |
|---|---|---|---|---|---|---|---|---|---|---|
| 1 | N | N | N/N | N/9th | 9th–25th/75th | N | N | N | Depression | N |
| 2 | N | N | N | N | 50th/99th | N | N | N | ADHD | N |
| 3 | 35 weeks/ NICU: 10 days poor feeding | N | N/N | Y/ < 0.4th | < 0.4th/2nd | High palate; dental overcrowding | Delayed and slow saccades | N | N | Choreoathetoid movements; precocious puberty (8 years) |
| 4 | 34 weeks/ Twin, IUGR; NICU: 3 weeks | N | N/N | N | < 0.4th/0.4th | N | End gaze nystagmus | N | N | Precocious puberty (9 years) |
| 5 | 34 weeks/ Twin, IUGR; NICU: 3 weeks | N | N/N | N | < 0.4th/0.4th | N | End gaze nystagmus | N | N | Precocious puberty (9 years) |
| 6 | Term/Need oxygen for 1 day | N | Mild/N | N/ | 50th/25th | Subtle dysmorphic features | N | N | Anxiety | Bicuspid aortic valve; 2 febrile convulsions |
| 7 | Mild feeding difficulties | N | Mild/N | N | 75th/50th/ | N | N | N | N | LL spasticity; facial hypomimia |
| 8 | Term/MAS | Y/speech delay | N/N | N/9th | NR/Short, no centile available | Elongated face; 5th finger clinodactyly | Myopic | N | N | IDDM; hypothyroidism |
| 9 | Term/IUGR | Y | Mild-Mod/N | Y/2nd | < 0.4th/ < 0.4th | Retrognathia | Slow vertical saccades | N | Emotional lability | Pyramidal signs |
| 10 | Term/N | N | Mild-Mod/N | Y/ < 0.4th | NR | Elongated face; bilateral 5th finger clinodactyly | N | N | N | Precocious puberty; pyramidal signs; GORD |
| 11 | Term/N | N | N/N | Y/ < 0.4th | 0.4th/0.4th | Subtle facial dysmorphic features | Oculomotor apraxia | N | N | N |
| 12 | Term/N | N | N/Y | Y/2nd | 50th/9th–25th | N | N | N | ADHD | N |
| 13 | Term/N | Y/speech delay | Mild/N | NR | 9–25th/0.4th–2nd | N | Slow saccades | Thickened toenails | Anxiety | Migraine |
| 14 | Term/IUGR | N | Mild/N | Y/ < 0.4th | 2nd/25th | Micrognathia, thick upper and lower lip; everted lower lip | Terminal nystagmus | N | N | Right-sided myoclonus; hyporeflexia |
| 15 | Term/N | N | N/N | N | 75th/9th | N | Astigmatism | N | N | N |
| 16 | Term/N | Y | Mild/N | N | NR/2nd | N | N | N | Depression; aggressive (occasional) | N |
| 17 | Term/N | Y/N | Mild/N | NR | NR/NR | Brachydactyly | Slow vertical saccades | N | Y; psychotic symptoms | Pyramidal signs |
| 18 | Term/N | N | N/N | NR | NR/NR | Bulbous nasal tip | N | N | N | N |
| 19 | Term/N | N | N/N | Y/ < 3rd | < 0.4th/NR | Bulbous nasal tip | NR | Dermatitis | N | N |
| 20 | Term/N | Y/speech delay | N/N | NR | 2nd/ < 0.4th | N | NR | N | N | Precocious puberty |

Patients 1–20 are shown. Patients 21–44 are provided in the Supplementary material. ADHD = attention deficit hyperactivity disorder; ASD = autistic spectrum disorder; Derm = dermatological; GORD = gastro oesophageal reflux disease; ID = intellectual disability; IDDM = insulin dependent diabetes mellitus; IUGR = intrauterine growth restriction; LL = lower limb; MAS = meconium aspiration syndrome; mod = moderate; N = no; NICU = neonatal intensive care unit; NR = not recorded; OFC = occipitofrontal circumference; Y = yes.

40), previous left middle cerebral artery infarct (Patient 30) and cerebellar vermis hypoplasia (Patient 41) (Table 1 and Supplementary material).

### Non-dystonic neurodevelopmental group

Within the cohort, we identified nine patients (seven females) with pathogenic mutations in *KMT2B*, in whom there has been no evolution of dystonia. Current median age is 11.8 years (range: 2.2–57.0 years). All presented with neurodevelopmental delay, with ensuing intellectual disability, microcephaly, short stature, and dysmorphic features (clinodactyly, syndactyly, facial dysmorphism) (Table 3). Additional systemic features include early feeding issues and intrauterine growth restriction. MRI brain was reported as normal in all patients in whom neuroimaging was performed ($n = 5$) with no evidence of GP hypointensity.

## Deep brain stimulation cohort

### Clinical features

Focused data on DBS outcome in *KTM2B*-dystonia was available for 18 patients (15 females), DBS was inserted for the management of medically refractory generalized dystonia. Median age at the time of the reporting was 14.5 years (range: 5.0–44.0 years), median age of onset of dystonia was 3.25 years (range: 2.0–10.0 years), and median time to generalization in this group was 2.75 years (range: 0.5–9.0 years). Evolution of dystonia culminating in status dystonicus before DBS surgery occurred in 4/18 subjects (22.5%). Lower limb dystonia was the presenting symptom in 16/18 subjects and laryngeal dystonia was a consistent finding with aphonia in 14. In 16/18 feeding was impaired, severely in six, requiring enteral nutrition. All patients developed independent ambulation prior to dystonia onset with progressive deterioration over time (Supplementary Fig. 2). Brain FDG-PET scan was performed in 12 patients; on visual reading, it was deemed normal in eight patients, with reduced basal ganglia uptake in three and heterogeneous cortical uptake in a single case (Patient 37). DaTSCAN performed in four patients did not provide evidence of dopaminergic neurodegeneration.

### DBS insertion and outcome

All patients received DBS to the GPi (Supplementary Table 10). Median age at DBS implant was 11.5 years (range: 4.5–37.0 years) and median duration from dystonia onset to surgery was 5.5 years (range: 2.0–35.0 years). Median postoperative follow-up was 2.0 years (range: 0.25–22.0 years) (Table 4). There were significant differences in BFMDRS-M scores between the preoperative, 6 months ($P = 0.001$) and 12 months ($P = 0.004$) groups. Significant changes were measured in BFMDRS-D between the preoperative, 6 months ($P = 0.009$) and 12 months ($P = 0.002$) assessments. Comparisons of the BFMDRS scores at later stages of follow-up (long-term subgroup, $>5$ years post-DBS, $n = 8$) confirmed maintenance of improvement for both compared to preoperative assessment, BFMDRS-M ($P = 0.012$) and BFMDRS-D scores ($P = 0.012$) (Table 4 and Fig. 3).

At 1 year, 8/15 cases assessed with the BFMDRS-M fulfilled criteria for an optimal response ($>30\%$). In the long-term subgroup, 5/8 cases sustained this improvement. For the BFMDRS-D scores, 7/15 showed improvement $>30\%$ at 1 year, maintained in 3/8 in the long-term subgroup. At the last assessment (median time 7.5 years, range: 5.0–22.0 years), dystonia improvement was maintained for the trunk (53.2%), neck (50.5%) and oro-mandibular (35.7%) regions. Improvement of dystonia in the lower limbs (16.3%) was inferior to the threshold set for responsiveness (Fig. 3E). At the last assessment, BFMDRS-D scores were variable, improvement maintained for swallowing (52.9%), dressing (40.0%) and writing (40.0%), whilst benefit on gait (16.2%) and speech (3.4%) were suboptimal (Fig. 3F). None of the eight subjects from the long-term subgroup maintained a fully autonomous gait; however, at initial stages of the therapy, three patients were able to ambulate without assistance (Supplementary Video 1, sequences 1 and 2). Freezing of gait during follow-up with DBS was documented in 5/8 patients, present in one before DBS (Patient 26) and during the DBS follow-up for the others (Supplementary Video 1, sequences 3 and 4). Four patients received DBS for status dystonicus (Supplementary Video 1, sequence 1). Patient 32 received initial GPi DBS with resolution of prolonged status dystonicus which had necessitated ICU management for 93 days. Six years after the first surgery additional leads to the subthalamic nucleus were implanted for a second episode of status dystonicus with a laryngeal component.

Eight hardware-related complications (five patients) were recorded. Patient 10 underwent surgical scar revision and bilateral extension cable replacement, electrode replacement occurred in two cases (Patients 9 and 21) due to worsening dystonia and high impedance and two cases had electrode revision due to a migrated electrode (Patients 20 and 55).

### Effect of genotype and severity of dystonia on response to DBS

Protein-truncating variants were identified in 11 subjects and chromosomal microdeletions in six. A single missense variant was identified and therefore not included in analysis (Table 1, Supplementary Table 7 and Supplementary material). No correlation was found between disease duration and dystonia severity preoperatively ($P = 0.257$). Evolution of BFMDRS-M scores as mutation type are presented in Fig. 4A. Similar preoperative BFMDRS-M scores were measured for protein-truncating variants (84.7) and chromosomal microdeletions (79) (Fig. 4B). Clinical response recorded at 1-year follow-up was superior for protein-truncating variants compared to chromosomal microdeletions (mean BFMDRS-M 50.4 versus 62.2). The XgBoost tree predictor importance model was used to define features important to predict mutation type. The most important features were the BFMDRS-M score preoperative (F-score, 38), 1 year (F-score, 21) followed by 6 months (F-score, 13) (Fig. 4C). No correlation was observed between

**Table 3 Phenotypic characteristics of patients with *KMT2B*-related non-dystonia phenotype ($n$ = 9)**

| Pt | Variant inheritance; how diagnosed | Age, y: Sex | Presentation reason | Birth/neo-natal Issues | DD/ID/ASD | Microcephaly/OFC (centile) | Weight/height (centile) | Dysmorphism | Eyes | Psychiatric | Other |
|---|---|---|---|---|---|---|---|---|---|---|---|
| 45 | c.1127_1130delAGGA p.Lys376Argfs*10 *de novo*; diagnostic WES | 4: M | GDD, feeding difficulties | 36 weeks/ DCDA twin; feeding issues | Y/Y/N | Y/ < 0.4th | 25th/9th–25th | Epicanthic folds; posteriorly rotated ears; bil. 5th finger clinodactyly | N | N | Hypotonia |
| 46 | c.3147_3160del p.Gly1050Profs*33 Unknown; identified in son (Pt17); research SS | 57: F | Son (Patient 17) with dystonia | NR | NR/Y/N | NR | NR/0.4th–2nd | Bulbous nose | N | N | N |
| 47 | c.3325delC p.Arg1109Glufs*73 Unknown; Identified in sibling (Pt 18); diagnostic panel | 29: F | ID | Term | Y/Y | NR | 91st–98th/ < 0.4th | Small hand and feet | N | N | Fatty liver - improved with dietary; congenital septal defect |
| 48 | c.3665G > A p.Cys1222Tyr *de novo*; research WES | 10.8: M | ID, dysmorphism | Term/feeding issues | Y/Mod/N | Y/2nd | 9th/9th | CDLS like craniofacial dysmorphism; ptosis; 2nd/3rd toe syndactyly | N | N | Cyclical vomiting; recurrent infection |
| 49 | c.3885 G > A p.Trp1295* *de novo*; diagnostic WES | 11.8: F | GDD | Term/GORD. poor weight gain | Y/Y/Y | N/50th | 25th–50th/ 25th–50th | Bil. 5th finger clinodactyly | CVI; astigmatism; hyperopia | ADHD | Stereotypies; constipation; urinary incontinence |
| 50 | c.4789C > T p.Arg1597Trp *de novo*; research WES | 12.3: F | Sev ID, partial disaccharide deficiency | IUGR; talipes; feeding issues | Y, speech delay/ Sev/N | Small for age, no centiles | 25th/25th | Abnormal palmar crease; epicanthic folds; straight nose | Strabismus | Challenging behaviour | Faecal incontinence; iron def anaemia |
| 51 | c.5046C > G p.Cys1682Trp *de novo*; diagnostic WES | 6.7: F | Speech delay | Term | Y, speech delay/N/ N | Yes | 2nd/25th | Bil. 5th finger clinodactyly; epicanthic folds; 2nd/ 3rd toe syndactyly | N | ADHD | N |
| 52 | c.6895dup p.Arg2299fs*4 *de novo*; diagnostic WES | 20: F | Feeding issues, ID | Feeding issues/ IUGR | Y, speech delay/ Mild/Y | Y/ < 0.2th | 2nd–9th/9th | Bil. 5th finger clinodactyly, syndactyly of 2nd and 3rd toes and clinodactyly of 4th and 5th toes Full cheeks, deep-set eyes with periorbital fullness | N | N | N |
| 53 | c.7271dupT p.Ser2425Glnfs*3 *de novo*; diagnostic WES | 2.2: F | Developmental delay | IUGR, RDS, Feeding issues; FTT | Y, global/Sev/N | Y/ < 3rd | 0.4th/9th | Frontal bossing, retrognathia, high arched uvula and inverted nipples; Bil. 5th clinodactyly | N | N | GORD, needing Nissen's fundoplication; hypotonia |

NM_014727.2, GRCh38 (Chr 19). ADHD = attention deficit hyperactivity disorder; ASD = autism spectrum disorders; Bil. = bilateral; CDLs = Cornelia de Lange syndrome; CVI = cortical visual impairment; DCDA = dichorionic diamniotic; DD = developmental delay; FTT = failure to thrive; GDD = global developmental delay; GORD = gastro-oesophageal reflux disease; ID = intellectual disability; IUGR = intrauterine growth restriction = Mod = moderate; RDS = respiratory distress syndrome; Sev = severe.

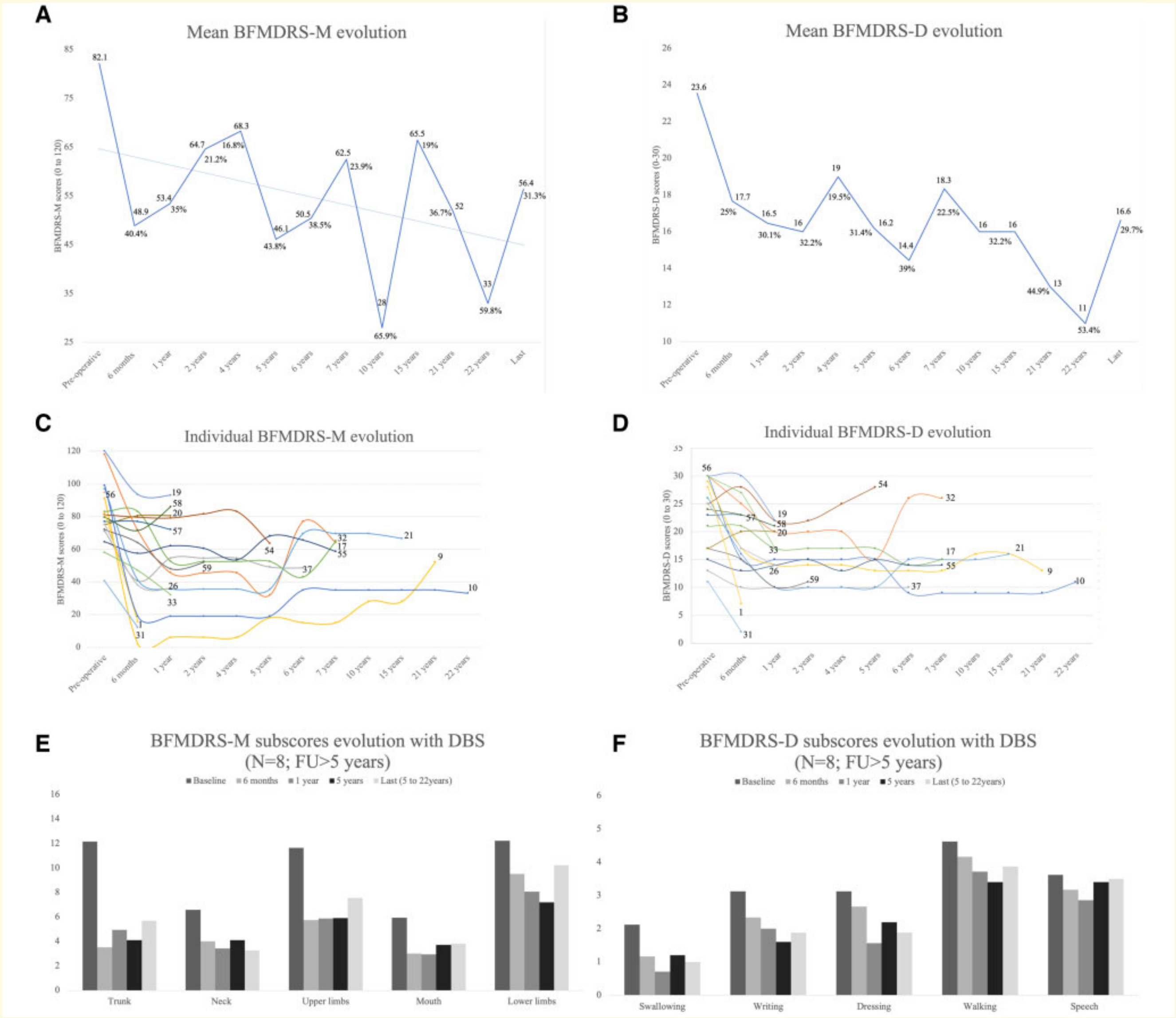

**Figure 3 Evolution of BFMDRS for the DBS cohort ($n$ = 18) and evolution of dystonia and motor function with DBS for patients followed > 5 years with DBS ($n$ = 8).** (**A**) Mean score of BFMDRS-M (range: 0–120). (**B**) Mean score of BFMDRS-D (range: 0–30). (**C**) Spaghetti plots displaying individual dystonia evolution with DBS, as assessed by the BFMDRS-M. (**D**) Spaghetti plots displaying individual dystonia evolution with DBS, as assessed by the BFMDRS-D. (**E**) Evolution of motor function as assessed by the motor section of the BFMDRS followed for > 5 years with DBS. (**F**) Evolution of motor function as assessed by the disability section of the BFMDRS followed for > 5 years with DBS.

DBS settings, disability score and at the last follow-up scores, for the long-term subgroup.

## Extended *KMT2B* cohort analysis

Overall, we identified 142 patients with *KMT2B*-related disease. Two patients with missense variants of uncertain significance in *KMT2B* were not included in analysis (Carecchio *et al.*, 2019; Ma *et al.*, 2019). Two individuals were duplicated in published papers (Lange *et al.*, 2017; Meyer *et al.*, 2017; Reuter *et al.*, 2017; Dafsari *et al.*, 2019). One paper was in Chinese and only the abstract was available in English for review (Dai *et al.*, 2019*b*). Three individuals included in our study cohort were also briefly reported in a cohort WGS study (Kumar *et al.*, 2019). Overall, 133 patients were included for analysis in the extended cohort. This comprises the patients reported in this study ($n$ = 53, study cohort), and a further 80 *KMT2B* mutation-positive cases reported in the literature (published cohort) (Supplementary Tables 4–6).

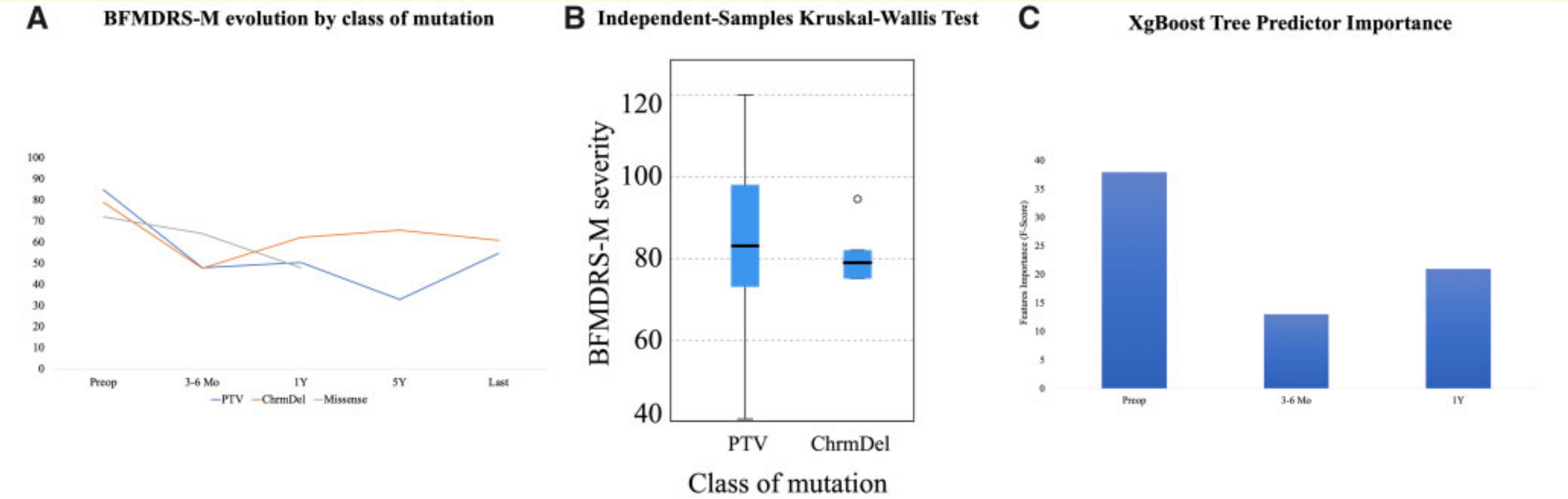

**Figure 4 Relationship between genotype, dystonia severity and DBS (*n* = 18).** (**A**) Dystonia BFMDRS-M scores evolution with DBS according to the class of mutation. Eleven protein-truncating variants (blue), six microdeletions (orange) and a single missense variant (grey). (**B**) Relationship between genotype and dystonia severity at baseline (BFMDRS-M). (**C**) XgBoost Tree Predictor Importance model was used to predict the type of mutation class according to the evolution of the motor scores. The BFMDRS-M scores preoperative (F-score, 38), 1-year post operative (F-score, 21) and 6 months (F-score, 13) were able to predict the type of mutation with 94.4% accuracy.

Within the extended cohort, 100 different intragenic mutations and 18 chromosomal microdeletions have been reported in 133 patients. Overall, protein-truncating variants are most frequently reported, accounting for 80 cases (60.2%), while chromosomal microdeletions (18 cases, 13.5%) and missense changes (35 cases, 26.3%) are less frequently described. Where segregation studies have been possible, the majority of mutations have either occurred apparently *de novo* (88.0%) or inherited from symptomatic parents (7.4%). Only five (4.6%) of reported cases harbour mutations that are inherited from an asymptomatic parent.

Of the 133 cases analysed, 123 (92.5%) cases present with a *KTM2B*-related dystonia phenotype (Supplementary Tables 5 and 6) (Zech *et al.*, 2016, 2017*a*, *b*; Lange *et al.*, 2017; Meyer *et al.*, 2017; Reuter *et al.*, 2017; Baizabal-Carvallo and Alonso-Juarez, 2018; Faundes *et al.*, 2018; Hackenberg *et al.*, 2018; Kawarai *et al.*, 2018; Zhao *et al.*, 2018; Brás *et al.*, 2019; Carecchio *et al.*, 2019; Dafsari *et al.*, 2019; Dai *et al.*, 2019*a*; Klein *et al.*, 2019; Kumar *et al.*, 2019; Ma *et al.*, 2019; Zhou *et al.*, 2019; Cao *et al.*, 2020; Mun *et al.*, 2020) The median age of dystonia onset is 5.0 years (range: 0.20–43.0 years), significantly lower in chromosomal microdeletions and protein-truncating variants (5.0 ± 3.8 years), compared to those with missense variants (6.0 ± 4.0 years, *P*-value = 0.0204) (Supplementary Fig. 4).

Systemic features (microcephaly, pre-existing developmental delay, intellectual disability) and hypointensity of GP on neuroimaging are more commonly described in patients with chromosomal microdeletions and protein-truncating variants than those with missense variants.

Within the published cohort, eight publications reported a total of 31 subjects treated with DBS for *KMT2B*-dystonia (Zech *et al.*, 2016, 2017*a*; Meyer *et al.*, 2017; Kawarai *et al.*, 2018; Carecchio *et al.*, 2019; Kumar *et al.*, 2019; Cao *et al.*, 2020; Miyata *et al.*, 2020; Mun *et al.*, 2020) Overall, results were expressed mostly as 'good or very good responses' and BFMDRS-M score changes were reported in 13 cases (Supplementary Table 8).

# Discussion

In this study, we describe the clinical and genetic features of the largest cohort of patients reported to date with *KMT2B* mutations, thereby elucidating a number of new and important concepts for this recently identified genetic disorder. Through detailed clinical delineation, we report subgroups of individuals with atypical dystonia presentations and non-dystonic phenotypes. In addition, we report the long-term outcome with DBS in 18 patients with *KMT2B*-dystonia, the largest cohort reported to date. This work has also identified clinically relevant genotype-phenotype correlations and provided deeper insight into the mutation spectrum of *KMT2B*-related disease and valuable data for DBS prognostication.

The majority of patients had either chromosomal microdeletions or protein-truncating variants, which have all either occurred *de novo* or with a fully penetrant autosomal dominant inheritance pattern. In contrast, missense mutations are less frequently reported. Where familial segregation studies have been possible, 88.0% occur *de novo*, 7.4% inherited from an affected parent and 4.6% from an apparently asymptomatic parent. The overall penetrance for *KMT2B*-related disease is therefore high, estimated to be 96.4%, with almost complete penetrance for protein-truncating variants and chromosomal deletions, and reduced penetrance (85.3%) for missense variants. These penetrance rates may still be an underestimate, as carrier parents may report no symptoms but be mildly affected with subtle sub-clinical disease features.

**Table 4 Dystonia evolution with GPI-DBS in the DBS cohort (*n* = 18)**

| Patient | Clinical information | | | | | BFMDRS evolution | | | | | | | | | |
|---|---|---|---|---|---|---|---|---|---|---|---|---|---|---|---|
| | Age DBS, y | Follow-up post-DBS, y | Target | SD post-DBS | Freezing of gait post-DBS | BFMDRS-M | | | | | BFMDRS-D | | | | |
| | | | | | | Pre-DBS /120 | 6 mo | 1 y | 5 y | Last | Pre-DBS /30 | 6 mo | 1 y | 5 y | Last |
| 1 | 4.5 | 0.5 | Bi-GPi | No | No | 82 | 15.5 | NA | NA | NA | 28 | 7 | NA | NA | NA |
| 9 | 5 | 21 | Bi-GPi | No | Yes | 91 | 2 | 6 | 18 | 52 | 29 | 17 | 14 | 13 | 13 |
| 10 | 8 | 22 | Bi-GPi | No | Yes | 99 | 19 | 19 | NA | 33 | 30 | 15 | 15 | NA | 11 |
| 17 | 23 | 7.5 | Bi-GPi | No | No | 83 | NA | 52.5 | NA | 65 | 21 | NA | 17 | NA | 15 |
| 19 | 7 | 1 | Bi-GPi | No | No | 120 | 93.5 | 93 | NA | NA | 30 | 30 | 22 | NA | NA |
| 20 | 7 | 1.5 | Bi-GPi | No | No | 75 | 80.5 | 80.5 | NA | NA | 17 | 20 | 20 | NA | NA |
| 21 | 28 | 15.5 | Bi-GPi | No | No | 97 | 41 | 35.5 | NA | 66.5 | 26 | 16 | 10 | NA | 16 |
| 26 | 15.5 | 1 | Bi-GPi | No | Yes | 75 | 38 | 36 | NA | NA | 25 | 17 | 14 | NA | NA |
| 31 | 15.5 | 0.25 | Bi-GPi | No | No | 40.5 | 12 | NA | NA | NA | 11 | 2 | NA | NA | NA |
| 32 | 7 | 7 | Bi-GPi | Yes | No | 118 | 71.5 | 45.5 | 32 | 64 | 30 | 25 | 20 | 15 | 26 |
| 33 | 11 | 1 | Bi-GPi | No | No | 58 | 47 | 32 | NA | NA | 30 | 27 | 17 | NA | NA |
| 37 | 37 | 6.5 | Bi-GPi | No | Yes | 71 | 40.5 | 54.5 | 49 | 48.5 | 13 | 10 | 10 | 10 | 10 |
| 54 | 7 | 5 | Bi-GPi | No | Yes | 81 | 79.5 | 79 | 63.5 | 63.5 | 25 | 28 | 22 | 28 | 28 |
| 55 | 12 | 8 | Bi-GPi | No | No | 64.5 | 57.5 | 62 | 68 | 58.5 | 15 | 13 | 14 | 15 | 14 |
| 56 | 10.5 | 0.1 | Bi-GPi | No | No | 94.5 | NA | NA | NA | NA | 30 | NA | NA | NA | NA |
| 57 | 10 | 1.5 | Bi-GPi | No | No | 77 | NA | 72 | NA | NA | 23 | NA | 21 | NA | NA |
| 58 | 4 | 1 | Bi-GPi | No | No | 79.5 | 71.5 | 86 | NA | NA | 24 | 23 | 21 | NA | NA |
| 59 | 8 | 2.5 | Bi-GPi | No | No | 72 | 64 | 48 | NA | NA | 17 | 15 | 10 | NA | NA |

Bi-GPi = bilateral GPi; BFMDRS-D/M = disability/motor section of the BFMDRS; mo = months; NA = not applicable; SD = status dystonicus; y = year.

Given the observed variable disease penetrance and paucity of functional diagnostic assays to assess KMT2B function, interpreting the clinical relevance of missense variants in *KMT2B* can often be challenging. In our study cohort, our threshold for deeming missense substitutions as potentially pathogenic has been based on a CADD score $>20$, $>2$ corroborative *in silico* predictions, absence from the gnomAD database and use of ACMG guidelines for determining variant pathogenicity (Supplementary Table 3) (Richards *et al.*, 2015). Using constraint analysis, we have demonstrated that protein-truncating variants are scattered throughout the entire gene, whereas missense variants that are thought to be pathogenic occur only in or around functionally important protein domains (Fig. 1). We advocate that all missense variants should be interpreted with caution, especially those occurring outside key domains, and/or with CADD scores $<20$.

Classical *KMT2B*-dystonia presents as an early childhood-onset progressive dystonia with prominent cervical, laryngeal and oromandibular involvement (Zech *et al.*, 2016; Meyer *et al.*, 2017). However, our study cohort also includes a subgroup of patients with atypical dystonia presentation at disease onset. Specifically, 7/44 cases initially presented with bulbar and laryngeal symptoms and only later developed limb involvement. Furthermore, 7/44 patients reported upper (rather than lower) limb involvement initially. This observation is further confirmed in our extended cohort analysis, where 16.4% of patients presented with an atypical dystonia phenotype, with features of oromandibular dystonia (dysarthria, change in quality or volume of voice) at first presentation. It is therefore increasingly evident that not all *KMT2B*-dystonia patients follow a typical course with caudocranial progression.

Our extended analysis of 133 cases has shown that the onset of dystonia appears to be significantly earlier in those with chromosomal microdeletions and protein-truncating variants than in those with missense variants (Supplementary Fig. 4 and Supplementary Table 5). The average age of dystonia onset seems earlier (5.0 years) when compared to other monogenic primary dystonia: 12 years in DYT-*TOR1A*, 14.0 years in DYT-*THAP1* and 31 years in DYT-*GNAL* (Blanchard *et al.*, 2011; Ozelius and Bressman, 2011; Fuchs *et al.*, 2013), this may be a clue to the underlying genetic diagnosis.

In our extended cohort analysis, the majority of patients with *KMT2B*-related dystonia (92.5%) have additional neurological, psychiatric and non-neurological systemic features, suggesting that most patients have a complex dystonia phenotype (Supplementary Tables 5 and 6). Many patients with *KMT2B* mutations present with an overlapping neurodevelopmental phenotype, and we propose that microcephaly, dysmorphism and intellectual disability should be recognized as core disease features. Mutations in other histone methylation modifier genes similarly cause phenotypically distinct neurodevelopmental syndromes, with these overlapping features, which are also reported in Wiedemann-Steiner syndrome (MIM: 605130, *KMT2A*), Kleefstra syndrome 2 (MIM: 617768, *KMT2C*), Kabuki syndrome 1 (MIM: 147920, *KMT2D*), Kleefstra syndrome 1 (MIM: 610253, *EHMT1*) and *SETD1A*-related disease (Ng *et al.*, 2010; Jones *et al.*, 2012; Kleefstra *et al.*, 2012; Singh *et al.*, 2016). Rodent models support a neurodevelopmental phenotype for *KMT2B*-related disease; conditional knockdown of *Kmt2b* in forebrain excitatory neurons leads to learning and memory impairment (Kerimoglu *et al.*, 2013). Our study thus further emphasises the key role of KMT2B in neurodevelopment.

Within our study cohort, we also identified a number of previously unreported features, including intrauterine growth restriction, early neonatal feeding issues and endocrinopathies (Table 2, Supplementary Table 6 and Supplementary material). Multiple endocrinopathies including growth hormone deficiency, pubertal disorders, and hypothyroidism are also reported in Kabuki syndrome, which is associated with defects in histone modification (Bereket *et al.*, 2001). The mechanisms underlying this derangement of endocrine function are not yet fully understood; proposed mechanisms include defective regulatory T cells or intrinsic B-cell tolerance breakage, both regulated by histone modification (Stagi *et al.*, 2016).

There are currently no validated biomarkers for *KMT2B*-related disease. The identification of neuroimaging abnormalities, in the context of a suggestive clinical phenotype, may facilitate diagnosis. In our study cohort, MRI features of bilateral GP hypointensity with a hypointense lateral streak of GP externus was reported in 56.2% of the overall cohort and in 83.9% of cases reviewed by our paediatric neuroradiologist (W.K.C.). This may be attributed to patient age at the time of neuroimaging, absence of SWI/B0 sequences, or neuroradiological expertise. The radiological signature does appear to be an age-dependent phenomenon, more likely to be present in younger patients (mean age at imaging, 11.0 years) than in older individuals (mean age at imaging, 23.6 years).

*KMT2B*-related dystonia appears refractory to commonly prescribed anti-dystonic agents. Although status dystonicus has only been previously described in two cases of *KMT2B*-dystonia (Meyer *et al.*, 2017; Cao *et al.*, 2020), 11.6% ($n = 5$) patients in our study cohort and 22.5% patients in the focused DBS cohort developed status dystonicus before DBS insertion. DYT-*KMT2B* appears to be one of the causes of dystonia with the highest risk of developing status dystonicus together with pantothenate-kinase associated neurodegeneration and *GNAO1*-related movement disorders compared to other monogenic childhood-onset dystonias such as DYT-*TOR1A* and *THAP1*-related dystonia which may be another helpful distinguishing disease feature (Opal *et al.*, 2002; Ben-Haim *et al.*, 2016; Koy *et al.*, 2018; Nerrant *et al.*, 2018; Oterdoom *et al.*, 2018; Waak *et al.*, 2018; Schirinzi *et al.*, 2019).

Fifty-two patients (23 in the study cohort and 29 patients in the published cohort) had GPi-DBS inserted for medically intractable dystonia (Tables 1, 4, Supplementary Table 8 and Supplementary material) (Coubes *et al.*, 1999; Zech *et al.*, 2016, 2017*a*; Meyer *et al.*, 2017; Kawarai *et al.*, 2018; Nerrant *et al.*, 2018; Zhao *et al.*, 2018; Carecchio *et al.*, 2019; Dafsari *et al.*, 2019; Miyata *et al.*, 2020; Mun *et al.*, 2020). In our focused DBS cohort, the median postoperative follow-up was 2.0 years, with the longest follow-up of 22.0 years. Dystonia was severe at the time of DBS surgery with a mean BFMDRS-M of 82.1 (120 being the severest dystonia score). Significant improvement was obtained for both mean BFMDRS-M and BFMDRS-D scores at 1-year post-DBS (35% and 30% reduction, respectively), improved or maintained at 5 years (44% and 31% reduction). At the last assessment, scores in the long-term subgroup ($n = 8$), showed sustained improvements of 31% and 29%, respectively. Dystonia improvement was maintained for trunk (> 50%), neck (> 50%) and oromandibular distribution (35.7%). Swallowing and upper limb function (dressing and writing) sustained a greater than 40% improvement compared to speech, which failed to change significantly at group level after DBS. Dystonia involving the lower limbs improved the least, despite an initial clinical improvement with the return of independent ambulation in some, worsening gait was documented in several patients (3/8) after DBS. No patient from the long-term subgroup maintained independent ambulation. Freezing of gait occurred post-DBS in five individuals (27.7%), more frequent than in other forms of monogenic dystonia (Schrader *et al.*, 2011). Under DBS, mild freezing of gait was observed the earliest at 3 years post-DBS (Patient 10) and documented at 6 years post-DBS (Patients 9 and 37). DaTSCAN was performed in two subjects with freezing of gait (Patients 9 and 37) and did not show striatal denervation. Comparing initial and long-term clinical outcomes in DYT-*KMT2B* with other types of monogenic dystonia, initial improvement is significant and comparable to outcomes observed in DYT-*THAP1* (Panov *et al.*, 2013; Danielsson *et al.*, 2019). However, as described in other forms of monogenic dystonia, secondary clinical worsening may occur, some patients becoming 'secondary non-responders'. Nonetheless, early age of dystonia onset, short mean time to generalization, pharmacoresistance and risk of status dystonicus should prompt early consideration for surgical management.

Our extended cohort analysis ($n = 133$) has revealed that patients with chromosomal microdeletions and protein-truncating variants have a higher burden of multi-system disease with microcephaly, developmental delay (before the onset of motor symptoms), intellectual disability, short stature and endocrinopathies, all of which are more frequently reported in this group than in those with missense variants (Supplementary Table 5). We detected a higher incidence of psychiatric features (such as anxiety, attention deficit hyperactivity disorder), microcephaly, low weight and short stature in our study group when compared to the published cohort; this observed difference may reflect the limitations of extrapolating information from published papers but may also suggest under-recognition of these associated features in *KMT2B*-dystonia (Supplementary Table 6). In the DBS cohort, preoperative BFMDRS-M dystonia scores appeared to be comparable in protein-truncating variants (84.7) and chromosomal microdeletions (79). Using the XGBoost Tree model, the features which predicted the type of variant were preoperative BFMDRS-M score, followed by 1-year and 6-month follow-up scores (accuracy 94.4%). There is an argument in favour of a relationship between the motor severity developed and the type of *KMT2B* variant. No correlation was found between DBS settings and the degree of clinical response.

Our study cohort has further confirmed that a small subgroup of patients with *KMT2B* mutations may not manifest dystonia (9/53, 17.0% cases). Despite this observed phenotypic pleiotropy, it is clear that both groups show a number of overlapping phenotypic features including microcephaly, dysmorphism, short stature, neonatal feeding issues, and early developmental delay evolving into intellectual disability. Within both our study cohort and the published cohort, it is conceivable that dystonia may not have yet developed in some patients and could potentially be a future disease feature, given that all the non-dystonia patients but one are under 30 years of age. Subtle features of dystonia (posturing, intermittent toe-walking) may not be appreciated by a non-movement disorder specialist. Moreover, given the relatively recent recognition of this *KMT2B*-subtype, the non-dystonia group may be under-recognized, and could account for a larger proportion of *KMT2B*-related disorders. The underlying disease mechanisms governing the manifestation of dystonia in typical disease (and absence of dystonia in the non-dystonia group and other *KMT2*-gene disorders) remain yet to be elucidated, but may be attributed to currently undetermined genetic, epigenetic and environmental factors. Our observations confirm that *KMT2B*-related disease represents a continuum from infancy to adulthood.

The mechanisms by which variants in *KMT2B* cause such a broad phenotypic disease spectrum and the reasons underpinning the observed genotype-phenotype correlations remain are yet to be fully elucidated. *KMT2B* encodes a histone lysine methyltransferase involved in methylation of the fourth lysine residue to histone 3 (H3K4). Although the exact function of this protein is not fully understood, it is thought to be a crucial regulatory mechanism for gene expression, active transcription and maintenance of genomic integrity, essential for the development and function of the CNS (Jenuwein and Allis, 2001; Kouzarides, 2007; Vallianatos and Iwase, 2015). It is postulated that haploinsufficiency or dysfunction of *KMT2B* affects the downstream expression of key genes regulating neurodevelopment and motor control. Knockout of *Kmt2b* in mice forebrain results in altered expression in the dorsal dentate gyrus of a number of genes associated with dystonia including *PRKRA* and *ADCY5* (Kerimoglu *et al.*, 2013). Other epigenetic and environmental factors may partially determine *KMT2B*-related phenotypes. Future work using patient-relevant cell and animal laboratory models of disease will assist in unravelling the underlying processes governing the *KMT2B* disease continuum.

# Acknowledgements

We thank all our patients and their families for taking part in this study. This research was supported by the NIHR Great Ormond Street Hospital Biomedical Research Centre. We also acknowledge support from the UK Department of Health via the National Institute for Health Research (NIHR) comprehensive Biomedical Research Centre award to Guy's and St. Thomas' National Health Service (NHS) Foundation Trust in partnership with King's College London. The research team acknowledges the support of the National Institute for Health Research, through the Comprehensive Clinical Research Network. The views expressed are those of the author(s) and not necessarily those of the NHS, the NIHR, Department of Health or Wellcome Trust. Sequencing for Patient 37 was provided by the University of Washington Center for Mendelian Genomics (UW-CMG) and was funded by the National Human Genome Research Institute and the National Heart, Lung and Blood Institute grant HG006493 to Drs Debbie Nickerson, Michael Bamshad, and Suzanne Leal.

# Funding

M.A.K. is funded by an NIHR Research Professorship and receives funding from the Sir Jules Thorn Award for Biomedical Research, Great Ormond Street Children's Hospital Charity (GOSHCC) and Rosetrees Trust. M.A.K., K.E.B., L.A., D.S., A.N., N.T. and E.M. are supported by the NIHR GOSH BRC. K.M.G. received funding from Temple Street Foundation. L.A. is funded by the Swiss National Foundation. E.M. received funding from the Rosetrees Trust (CD-A53), and the Great Ormond Street Hospital Children's Charity. A.S.J. is funded by NIHR Bioresource for Rare Diseases. S.A.I. and M.H. are supported by the NINDS Intramural program. K.P.B. is PI of the Movement disorders centre (MDC) at UCL, Institute of Neurology which has been funded by the BRC. He has grant support by EU Horizon 2020. M.E.D-H. has clinical training grant through Tourette Association of America, but the research is unrelated to *KMT2B*. T.L. received funding from Health Research Board, Ireland and Michael J Fox. Foundation. K.A.M. receives funding from the NIH (award number K23NS101096-01A1). N.S. receives funding from the NIH (award number NS 087997 0). D.D. was supported by KIM MUSE Biomarkers and Therapy study grant during this work. B.B.A.d.V. financially supported by grants from the Netherlands Organization for Health Research and Development (912-12-109). J.F. is funded by the Rady Children's Institute for Genomic Medicine. F.L.R. is funded by Cambridge Biomedical Research Centre. The DDD study presents independent research commissioned by the Health Innovation Challenge Fund [grant number HICF-1009-003], a parallel funding partnership between the Wellcome Trust and the Department of Health, and the Wellcome Trust Sanger Institute [grant number WT098051]. This research was made possible through access to the data and findings generated by the 100 000 Genomes Project (Patient 34). The 100 000 Genomes Project is managed by Genomics England Limited (a wholly owned company of the Department of Health). The 100 000 Genomes Project is funded by the National Institute for Health Research and NHS England. The Wellcome Trust, Cancer Research UK and the Medical Research Council have also funded research infrastructure. The 100 000 Genomes Project uses data provided by patients and collected by the National Health Service as part of their care and support. Research reported in this manuscript was supported by the NIH Common Fund, through the Office of Strategic Coordination/Office of the NIH Director to the Undiagnosed Disease Network (UDN) and the NIH Undiagnosed Disease Program (Award numbers: U01HG007690 and U01HG007703). The content is solely the responsibility of the authors and does not necessarily represent the official views of the National Institutes of Health.

# Competing interests

K.P.B. has received honoraria to speak in sponsored meetings or act as consultant on advisory boards from Ipsen, Allergan, Teva, Cavion, and Retrophin pharma companies. He receives book royalties from Oxford University Press and Cambridge university press. T.L. served on a Scientific Advisory Board for An2H in 2018. J.F. spouse is Founder and Principal of Friedman Bioventure, which holds a variety of publicly traded and private biotechnology interests. In addition, he is chief operating officer of DTX Pharma which is a company developing RNA therapeutics. All other authors report no competing interests.

# Supplementary material

Supplementary material is available at *Brain* online.